\documentclass{article}

\usepackage{arxiv}
\usepackage[utf8]{inputenc} % allow utf-8 input
\usepackage[T1]{fontenc}    % use 8-bit T1 fonts
\usepackage{hyperref}       % hyperlinks
\usepackage{url}            % simple URL typesetting
\usepackage{booktabs}       % professional-quality tables
\usepackage{amsfonts}       % blackboard math symbols
\usepackage{nicefrac}       % compact symbols for 1/2, etc.
\usepackage{microtype}      % microtypography
\usepackage{lipsum}		% Can be removed after putting your text content
\usepackage{graphicx}
\usepackage{natbib}
\usepackage{doi}
\usepackage{bm}
\usepackage{amsmath}
\usepackage{graphicx}
\usepackage[svgnames]{xcolor}
\usepackage{tcolorbox}

\title{The speckle contrast extended to the polarimetric case: applications to radar and Laser images}

\author{  Elise Colin \\
	DTIS-Onera, Palaiseau, France, F-91123, Université de Paris Saclay, \texttt{elise.colin@onera.fr} \\
}

\date{}
\begin{document}
\maketitle

\begin{abstract}
This article proposes the application of various alternative definitions of the multivariate coefficient of variation parameter in two domains: radar polarimetric time series and dynamic polarimetric speckle. In the first case, the focus is on detecting permanent scatterers or changes, while in the second case, it involves calculating activity images. Our study demonstrates that most of these parameters offer added value in terms of signal-to-noise ratio improvement and enhancing contrast in specific regions. Furthermore, the concept of polarimetric multivariate coefficient of variation proves to be closely related to the degree of polarization.
\end{abstract}

% keywords can be removed

\section{Introduction}

Coherent sensors, such as radars and lasers, offer unprecedented capabilities for observing and analyzing complex scenes in various domains, ranging from remote sensing to medicine. One distinctive characteristic of these sensors is the presence of "speckle," a phenomenon resulting from the constructive and destructive interference of waves reflected or backscattered by objects in the observed scene. However, instead of considering speckle as a limitation or problem, it can be harnessed beneficially by utilizing temporal series of speckle.

Temporal series of speckle are obtained by acquiring coherent images at multiple time instances. This approach captures the dynamic properties of speckle, which reveal valuable information about changes and motion in the observed scene. In the field of radar remote sensing, for example, analyzing temporal series of speckle has led to the development of advanced change detection algorithms. By monitoring temporal variations in speckle, it is possible to detect and characterize modifications in the environment, such as changes in land cover, agricultural monitoring, snow cover.

Similarly, in the field of lasers, exploiting temporal series of speckle offers exciting prospects for analyzing biological tissues. Using dynamic speckle techniques, it is possible to obtain maps of activities or motion within tissues. This approach has significant applications in medicine, enabling the detection of blood flow under the skin, in the brain, or tracking tumor growth, and in biology with the sap and senescence monitoring.

One of the key parameters utilized in this monitoring is the speckle contrast, which is defined as the ratio between the standard deviation and mean of the temporal signal. It is also commonly referred to as the coefficient of variation. The speckle contrast provides a measure of the temporal fluctuations in the speckle pattern, reflecting changes and motion within the observed scene.

In this paper, the novelty lies in investigating the various extensions of the speckle contrast parameter to polarimetric acquisitions. Polarimetry plays a crucial role in capturing additional information about the polarization properties of the scattered or reflected light. 

Previous research has demonstrated the crucial role of polarimetry in radar change detection. Indeed, certain changes can only be detected in a specific polarimetric channel. In the field of optics, the LSOCI technique utilizes orthogonal polarization to enhance perceptible activity images. This enhancement is based on the principle that a filter using orthogonal polarization enhances multiple scattering and improves the obtained activity maps.

This article seeks to enhance the formalism of speckle contrast in the context of polarimetric acquisitions by exploring proposed mathematical extensions. Currently, there exist four parameters that have been developed to extend the notion of coefficient of variation to the multimodal case. The objective of this study is to examine these mathematical extensions and determine their applicability in characterizing the polarimetric variations observed in coherent sensors. By expanding the understanding and utilization of these parameters, we aim to improve the quantitative analysis and interpretation of polarimetric speckle data.

In Section 2, we will present these four alternatives and discuss their application to our radar and laser acquisitions. We will then apply these concepts to the two reference domains of dynamic speckle imaging: radar satellite imagery in Section 4 and LSCI biomedical imaging in Section 5. The synthesis of the results obtained with these different approaches will be provided in Section 6. This comprehensive analysis aims to shed light on the effectiveness and suitability of these alternative approaches in improving the analysis and interpretation of polarimetric speckle data in both satellite remote sensing and biomedical imaging applications.

\section{Extansion to the coefficient of variation and applicability to the speckle contrast}

The speckle contrast is a quantitative measure that characterizes the level of spatial variation or granularity observed in a speckle pattern. Speckle patterns are typically encountered in coherent imaging systems, such as radar or laser imaging, where interference effects occur due to the interaction of waves with random scatterers or reflective surfaces.
The speckle contrast is calculated as the ratio of the standard deviation to the mean intensity of the speckle pattern.

The speckle contrast can be calculated either spatially, by considering a population of connected pixels within a neighborhood, or temporally, by considering the set of values taken by a given pixel over a time series.

However, from a mathematical standpoint rather than a physical one, the definition of speckle contrast is rigorously the same as that of the coefficient of variation.

Coefficient of variation is a statistical measure that quantifies the relative variability of a dataset or a distribution. It provides a standardized measure of dispersion, allowing for the comparison of variability between datasets with different means and scales.

Several studies have suggested an extension of this definition to accommodate multimodal samples.

Let $P$ be the multimodal time-series of dimension $N$. Each acquisition $\mathbf{p}^{i}$ is a real vector of dimension $p$. 
\begin{equation}
 \mathbf{P}=\begin{pmatrix}
\mathbf{p}^{1}, & ... & \mathbf{p}^{k} & ... & \mathbf{p}^{N} \end{pmatrix}
\end{equation}

We consider the four mathematical formulations existing for the corresponding multimodal coefficient of variation:

\begin{equation}
  \gamma_{\textrm{R}}=\sqrt{\frac{\det(\mathbf{C})^{1/p}}{\mathbf{\bm{\mu}^{\dag} \bm{\mu}}}}, ~~\gamma_{\textrm{VV}}=\sqrt{\frac{\textrm{trace}(\mathbf{C})}{\mathbf{\bm{\mu}^{\dag} \bm{\mu}}}},  ~~   \gamma_{\textrm{VN}}=\sqrt{\frac{1}{\mathbf{\bm{\mu}^{\dag} C^{-1}\bm{\mu}}}}, ~~\gamma_{\textrm{AZ}}=\sqrt{\frac{\mathbf{\bm{\mu}^{\dag} C \bm{\mu}}}{(\mathbf{\bm{\mu}^{\dag} \bm{\mu})^2}}}
\end{equation}

where $\bm{\mu}$ is the $p$-dimensional vector containing the $p$ temporal means, $\bm{C}$ is the temporally estimated coherence-covariance matrix of the vector $\mathbf{p}$. 

Note that these formulations are all invariant by unit basis change . Moreover, in these formulations, the components of the signals are real values. 

$\gamma_{\textrm{R}}$ proposed by \cite{ra7studies} was introduced in the 1960s. This parameter suffers from the drawback of rapidly approaching zero when any of the eigenvalues of the matrix $\textbf{C}$ is close to zero. The approach proposed by \cite{van2005variation} for $\gamma_{\textrm{VV}}$ does not have this drawback but does not consider correlations between dimensions. $\gamma_{\textrm{VN}}$ proposed in \cite{voinov2012unbiased} incorporates these correlations, but the result is often noisy as the definition involves matrix inversion of $\textbf{C}$.
The paper \cite{albert2010novel} introduced the last new formulation of the multivariate coefficient of variation $\gamma_{\textrm{AZ}}$. The paper aims to addressing certain limitations of the previous definitions, and enhancing its applicability in various fields of study. The novel formulation takes into account the correlations and interdependencies among the variables, providing a more comprehensive measure of their variability. The paper also provides mathematical derivations and analytical insights to support the proposed formulation. Moreover, practical applications and examples are provided to demonstrate the usefulness and effectiveness of the novel multivariate coefficient of variation in real-world scenarios.

\cite{aerts2015multivariate} focuses on comparing different definitions and properties of multivariate coefficients of variation, including their robustness to outliers, efficiency in estimating variability, and sensitivity to changes in the data distribution. Aerts evaluates and compares these measures using simulated data and real-world datasets, providing empirical evidence for their performance. The paper concludes emphasizing the importance of choosing the most suitable measure based on the specific requirements and characteristics of the data at hand, by taking into account factors such as data structure, dimensionality, and robustness considerations. In all cases, the samples provided were based on statistical data; however, to our knowledge, they have not been applied to images. For this reason, we aim to analyze their practical implementation on speckle images, first in the case of radar and then in the case of coherent laser images. In both cases, the multidimensional aspect arises from polarimetry.

\section{Radar time-series Applications}

The paper \cite{colin2020change} addresses the topic of change detection in urban areas using Synthetic Aperture Radar (SAR) time-series data. The paper focuses on urban environments, where changes can occur frequently due to urban development, construction activities, and land cover changes. 
The study demonstrates the potential of the coefficient of variation as a robust measure for change detection in urban areas using SAR time-series data. The authors validate their approach using real SAR time-series and simulation data, and compare the results of their change detection method with ground truth data, such as land cover maps or reference images, to evaluate its accuracy and effectiveness.

Radar polarimetric data refers to data obtained by measuring the complex electric field for various polarization combinations in transmission and reception. 
In radar time series, the coefficient of varition is used for two purposes: the detection of permanent scatterers \cite{colin2021urban}, and the detection of changes \cite{colin2020change}. In both cases, it has been shown that having a diversity of polarizations improves the detections. Indeed, deterministic objects often have a preferred orientation axis, and in this case, a preferred polarimetric response as well. In \cite{colin2019}, it was shown that over continental areas, only total polarimetry could detect a maximum of changes. The removal of a polarimetric channel is accompanied by the loss of at least 25 percent of detections. For this reason, in order to extend the algorithms that rely on the coefficient of variation to detect changes to a polarimetric scenario, the maximum of the coefficients of variation in each of the available polarimetric channels is classically used. Here, we go further, and propose the analysis of the different existing multivariate coefficients of variation.

It should be noted that the definitions of multivariate coefficients only apply to real vectors and cannot be extended to the complex case. Therefore, to incorporate these data into the aforementioned definitions, we propose two options:

- In the first approach, we focus solely on the magnitudes of the signals in each polarization channel.

- In the second approach, we convert our complex field signal into intensity data using the Stokes formalism.

Complete polarimetric time-series data is rare. Their equivalent in the Stokes formalism involves studying square 4x4 matrices with 16 coefficients. In this article, we have chosen to focus on the case study of time-series (VV, VH), that can be converted into a 4-component Stokes vector.

We applied these 4 formulations to several scenarios of TanDEM-X data over Toulouse (France) \footnote{provided by DLR under the scientific proposal OTHER0103. This work would not have been possible without the collaboration of Paola Rizzoli and Jose Luis Bueso Bello in DLR, whom we thank very sincerely. This fruitful collaboration was initiated in the framework of the ONERA-DLR virtual research center for cooperation in "AI and applications in aerospace engineering".}: a dual pol HH/VV case.
In the case where we consider the amplitudes of our signals, we have vectors of dimension p=2. If we convert this vector to dimension 4, we end up with p=4.

\begin{figure}
    \centering
    \includegraphics[width=17cm]{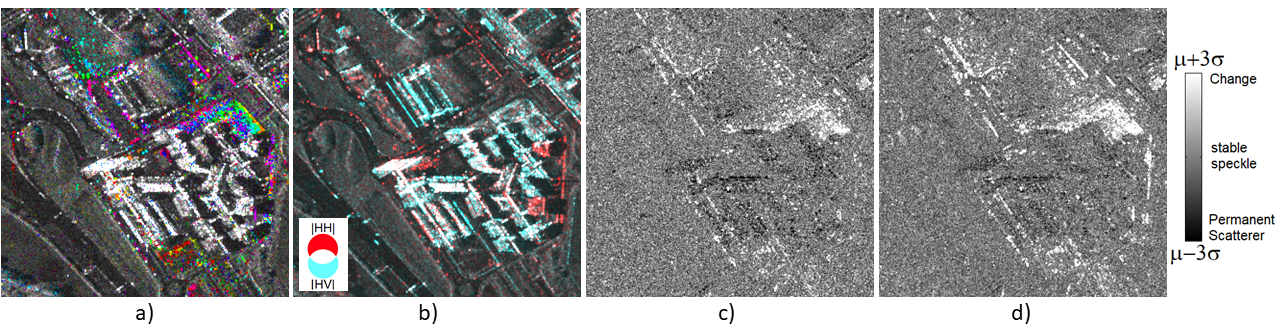}
    \caption{ a) the REACTIV representation highlights changes in colors, b) relative co-pol and crosspol channels, c) Coefficient of varition in Single Pol, d) Multivariate Coefficient of variation from \cite{albert2010novel}}
    \label{fig:my_label1}
\end{figure}

We analyzed the different coefficients obtained on a small area of the image that contains several buildings, some of which are under construction.
In Figure \ref{fig:my_label1}, the REACTIV representation on the left (a) reveals color-coded points subject to radiometric changes, with white points indicating "permanent scatterers."
In (b), we show the intensity differences obtained in the two polarimetric channels (HH) and (HV) using a color composition of two complementary colors. In (c), we displayed the single-channel contrast coefficient (HH), and in (d), the multivariate coefficient $\gamma_{AZ}$.

In both cases (c) and (d), we can observe the areas of high intensity changes in pixels, indicating zones of change, as well as areas of low intensity variations, indicating stable regions. 

\begin{figure}
    \centering
    \includegraphics[width=17cm]{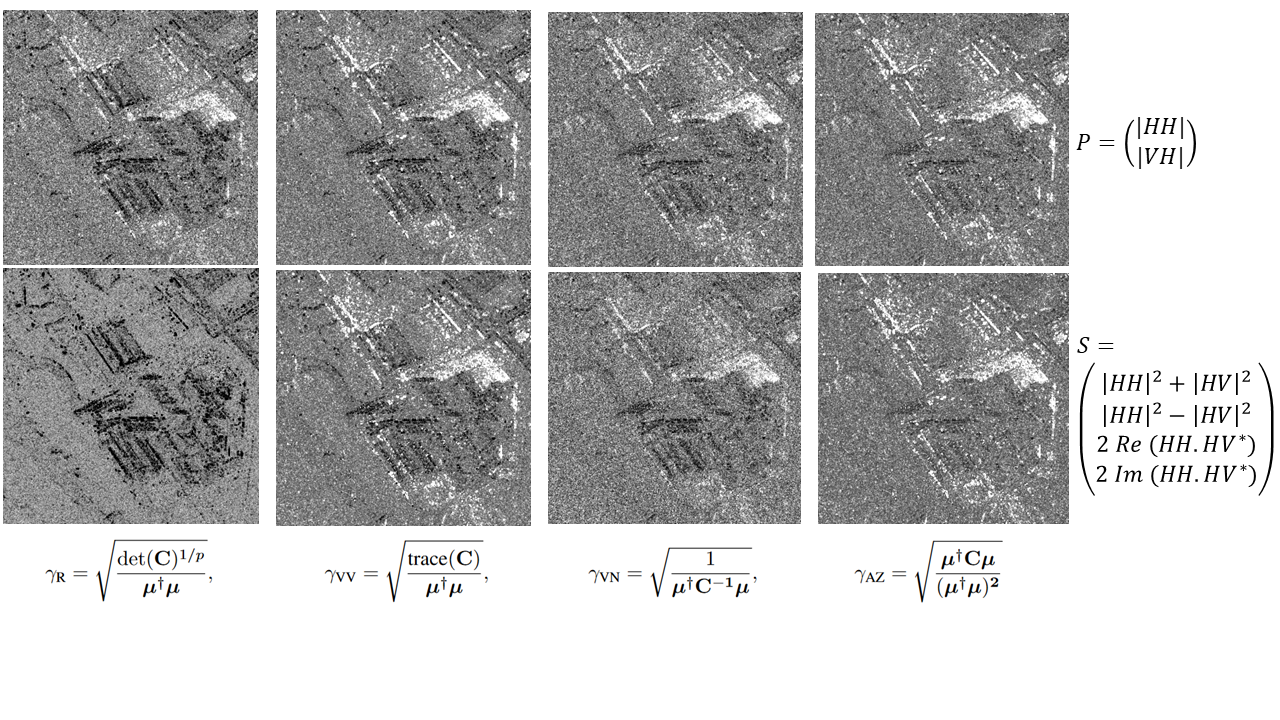}
    \caption{Comparison of the different multimodal contrast coefficients obtained in the case of p=2 (top) and p=4 (bottom) (taking into account cross-correlations between channels).}
    \label{fig:my_label2}
\end{figure}

Our observations in figure \ref{fig:my_label2} regarding the analysis of these coefficients of variation applied to the time series with p=2 or p=4 is as follows:

\begin{itemize}
\item The calculation of multimodal contrast coefficients leads to a less noisy signal in homogeneous areas, introducing a smoothing effect.

\item The coefficient $\gamma_R$ highlights Permanent Scatterers the best. This aligns with a mathematical property mentioned earlier: when one eigenvalue is close to zero, this coefficient also becomes very close to zero. On the other hand, the coefficient that finds the least Permanent Scatterers seems to be $\gamma_{AZ}$.

\item The three coefficients $\gamma_{AZ}$, $\gamma_{VV}$, and $\gamma_{VN}$ show little difference between them, both in the cases of p=2 and p=4. Therefore, considering the phases between the two polarimetric channels does not significantly alter the result.

\item Regarding $\gamma_R$, the difference is striking. It seems ill-suited for change detection: when the signal is stable in one of the polarimetric channels, potentially detectable differences in the other channels go unnoticed in the final calculation of the coefficient.

\end{itemize}

These initial observations will deserve further in-depth analysis in the future.

\section{Applications to Laser dynamic speckle}

Recently, a study in \cite{plyer2023imaging} was able to perform measurements of dynamic speckle using a polarimetric Sony camera on the vascularization of a beating heart.

The primary difficulty of the measurement was imaging a moving object, as the heart was placed in a perfusion machine that sustained its cardiac motion. To address this, a proposed solution involved extracting speckle images from the measurement series for closely positioned hearts and subsequently computing the corresponding contrast images.

Therefore, we proceeded with this type of measurement, considering a time-series of 40 speckle acquisitions for heart views with a closely matched geometric configuration. These time-series were acquired for four primary polarization directions: 90°, 135°, 0°, and 45° inclination with respect to the emission polarization direction.

We are interested in calculating the inverses of the squared contrast coefficients to obtain a homogeneous index comparable to the VMAI index defined in \cite{plyer2023imaging}. This index is expressed through the inverse of the contrast coefficient. It is physically related, under certain assumptions of physical models, to the inverse of the decorrelation time of the electric field. Thus, a high VMAI (Variation of Multimodal Amplitude Index) allows for the detection of motion.

\begin{figure}
    \centering
    \includegraphics[width=16cm]{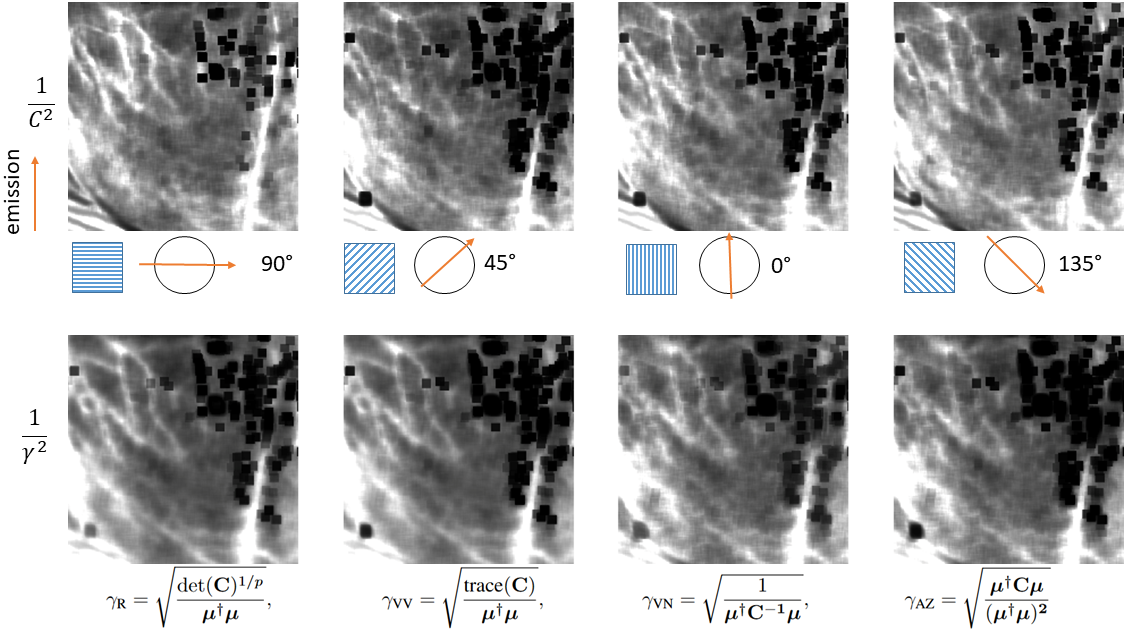}
    \caption{Images of VMAI calculated using single-channel and multivariate contrast coefficients, with spatial estimation.}
    \label{fig:optic1}
\end{figure}

In the first step, we calculate the contrast coefficients and the associated VMAI using purely spatial estimation on a single image from our sequence, on a 5x5 boxcar filtering window. The results are presented in Figure \ref{fig:optic1}.

In terms of spatial contrast calculation, the VMAI image at a 90° polarization angle exhibits the highest contrast compared to all others, even if the original intensity image having the lowest signal-to-noise ratio. The four multimodal contrast images show a similar trend with additional smoothing effects. However, this smoothing effect is not desirable when estimating spatial contrast since spatial estimation already introduces a loss of spatial resolution. Moreover, multimodal contrasts are more affected by signal saturation in specular regions, as these effects may only appear in one of the polarimetric channels but are reflected in all multimodal contrasts.

\begin{figure}
    \centering
    \includegraphics[width=16cm]{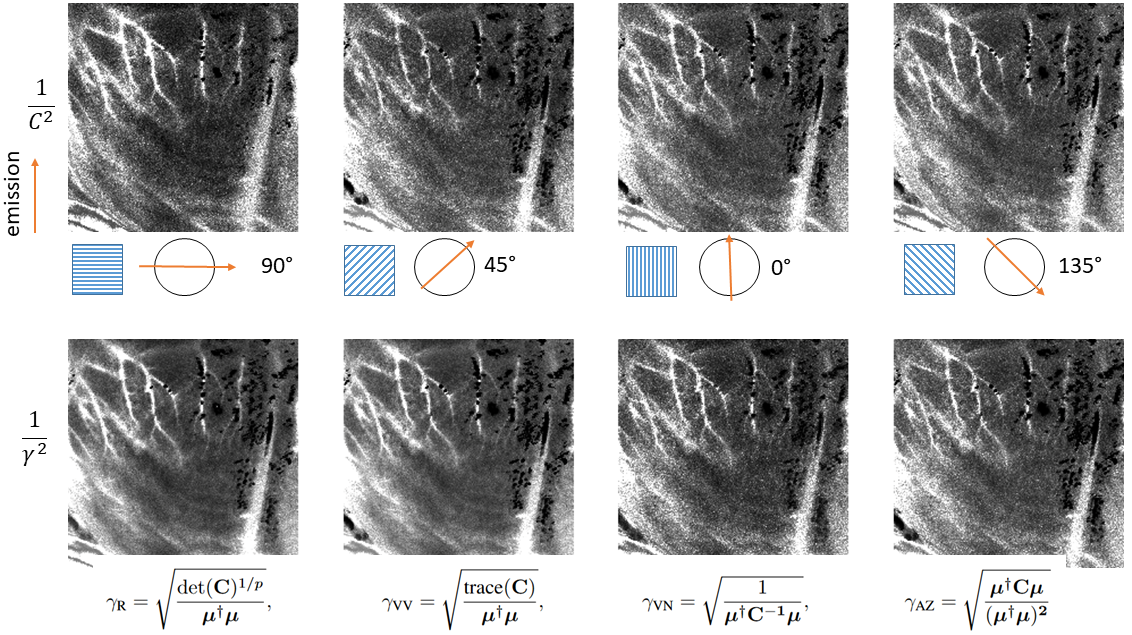}
    \caption{Images of VMAI calculated using single-channel and multivariate contrast coefficients, with temporal estimation.}
    \label{fig:optic2}
\end{figure}

For temporally calculated contrasts in figure \ref{fig:optic2}, once again, the monopolarimetric contrast performs the best in cross-polarization. The benefits of temporal coefficients are more pronounced, accompanied by improved signal-to-noise ratios between vessel and non-vessel regions.

\begin{figure}
    \centering
    \includegraphics[width=10cm]{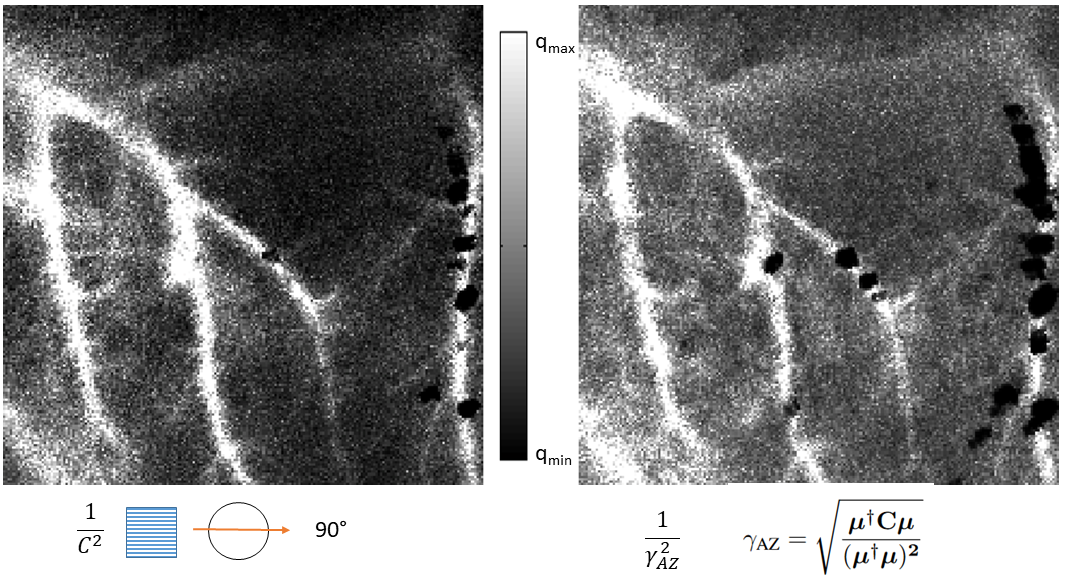}
    \caption{Left: VMAI image obtained from the contrast calculated in cross-polarization. Right:  VMAI image calculated using $\gamma_{AZ}$}
    \label{fig:optic4}
\end{figure}

\begin{figure}
    \centering
    \includegraphics[width=13cm]{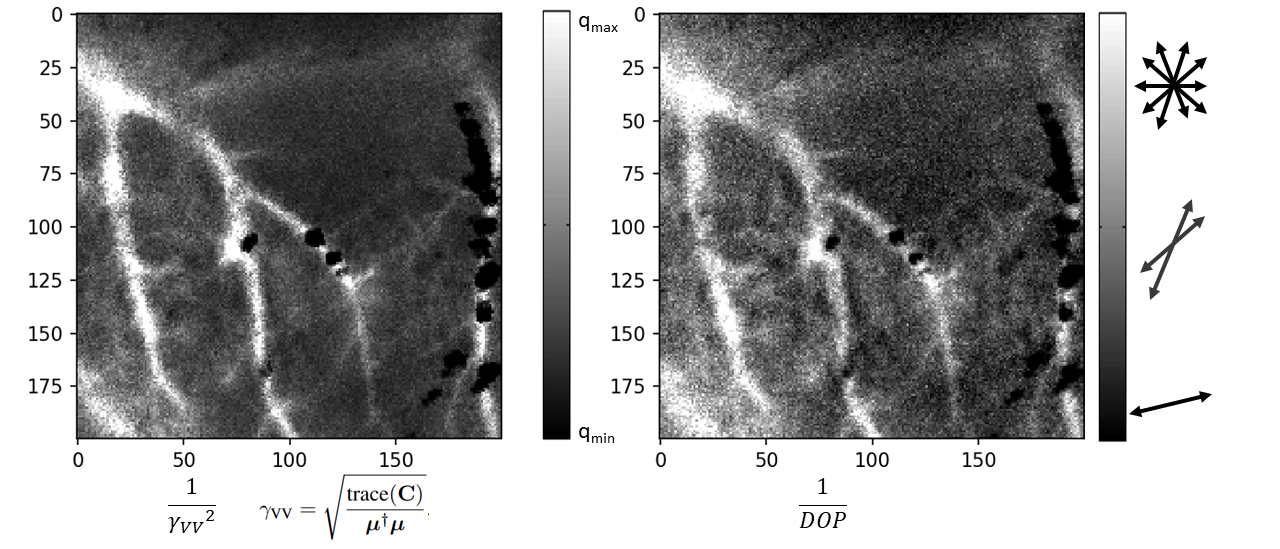}
    \caption{A comparison between the VMAI image obtained from the $gamma_{VV}$ coefficient and the depolarization image calculated as the inverse of the DOP (Degree of Polarization)}
    \label{fig:optic3}
\end{figure}

To better understand the effects on image details, we compare specific excerpts of the images. We compare the VMAI image obtained from the contrast calculated in cross-polarization with the VMAI image calculated from $\gamma_{AZ}$ on the figure \ref{fig:optic3}, and with the VMAI image calculated from $\gamma_R$ on the figure \ref{fig:optic4}.

\begin{figure}
    \centering
    \includegraphics[width=10cm]{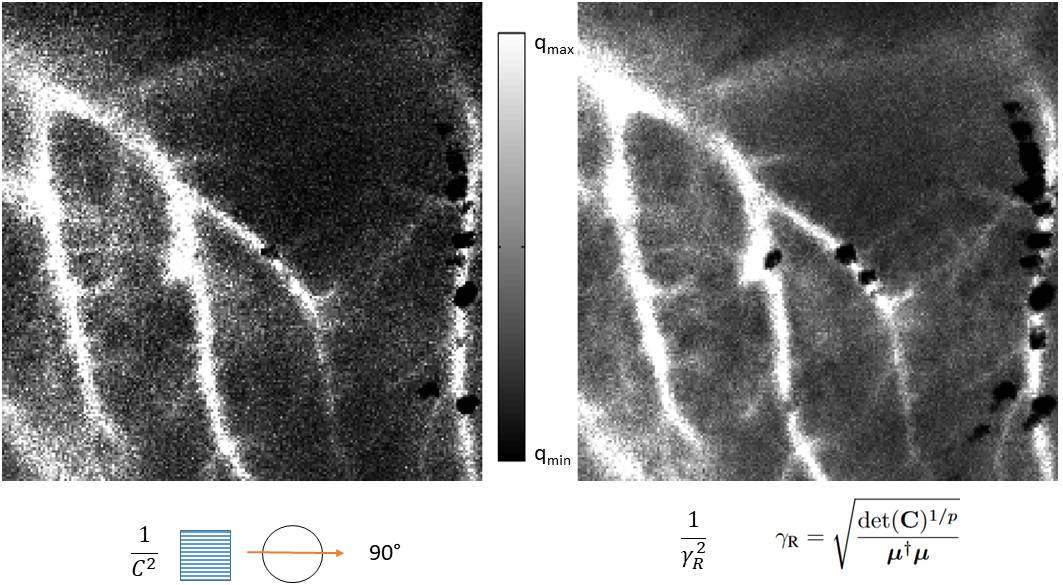}
    \caption{Left: VMAI image obtained from the contrast calculated in cross-polarization. Right:  VMAI image calculated using $\gamma_R$}
    \label{fig:optic5}
\end{figure}

The following trends are observed: 
- the contrast involving the inverse matrix $CV_{VN}$, is the most noisy among all, while the both least noisy contrasts are $CV_{R}$ and $CV_{VV}$. Certainly, the difference might be due to the fact that the formulations of $CV_{VN}$ and $CV_{AZ}$ involve cross-correlations between polarimetric channels.

- The images obtained from the multimodal coefficients capture new details in regions of intermediate intensity.

- However, they suffer from the drawback of amplifying signal saturation phenomena in specular reflection areas.

Finally, partial temporal degree of polarization was calculated. The resulting image closely resembles the multimodal contrast images, as we can observe in figure \ref{fig:optic5}

Correlation analysis using Pearson's coefficient reveals that the most correlated image is the $CV_{VV}$ coefficient with a correlation of 0.90, followed by $CV_{R}$ at 0.88, $CV_{AZ}$ at 0.80, and $CV_{VN}$ at 0.76.

Analyzing the least correlated areas of the image, we have seen that these areas include:

- Deeper vessel regions where the depolarization effect is slightly less visible compared to the blurring effect of contrast.
- Specular signal regions where the coefficient of variation cancels out, and the activity index saturates faster than depolarization.

Once again, these effects warrant further analysis on a other sample images.

\section{Conclusion}

In this article, we applied different possible definitions of a multimodal coefficient of variation to the analysis of polarimetric speckle time series. We conducted initial experiments in two application domains: radar imaging and imaging of blood flow activity in a heart.

In the first case, we had access to dual-pol data (HH and HV) which we evaluated, first on bimodal vectors in amplitude, and then on Stokes vectors.

In the second case, we analyzed time series acquired using a 4-channel polarimetric Sony camera on images of cardiac grafts.

We observed the following points:

\begin{itemize}
    \item  In laser imaging, the cross-channel provides the best speckle contrast images compared to other single-channel speckle contrast images.
    \item     Multimodal contrast coefficients tend to improve the signal-to-noise ratio in homogeneous areas, both for radar and optical domains.
    \item Multimodal contrast coefficients involving cross-correlations are more susceptible to noise, both for radar and optical domains.
    \item $\gamma_R$ reveals the Permanent Scatterers (low values) better but struggles to detect changes.
    \item 
    Temporal depolarization is highly correlated with multimodal coefficient of variation.
\end{itemize}

For this last observation, it will be necessary to first analyze the link between the mathematical formulations of depolarization and multivariate coefficient of variation, and then investigate the existence of media that depolarize without blurring the temporal contrast, or conversely, deterministic media that increase the temporal contrast.

\section*{Acknowledgements}

First and foremost, I would like to express my sincere gratitude to Professor Albert for graciously sharing his work with me.
I would also like to thank Dr. Aurélien Plyer, who worked alongside me in developing Python tools capable of processing our dynamic speckle data, both for radar using the Python module turtlesar and for optics using the pitae module.

These efforts are part of the MUSIC Chair project, which is funded by ONERA. 
The radar component also received support from the AI4GEO project funded by BPI. The optical laser component received support from the ECHORONEX project. ECHORONEX study was supported by Fondation Coeur et Recherche (Paris, France) and the Association pour le Développement et l’Amélioration des Techniques de Dépistage et de Traitement des Maladies Cardiovasculaires (ADETEC, Suresnes, France). I am deeply grateful to Professor Julien Guihaire for enabling me to carry out and analyze these acquisitions.

\bibliographystyle{unsrtnat}
\bibliography{sample} 

\end{document}